\documentclass[preprint,prl,showpacs,preprintnumbers,amsmath,amssymb]{revtex4}
\usepackage{amsfonts}
\usepackage{amssymb}
\usepackage{graphicx}

\begin{document}

\title{Search for the quark shell structure using the non-topological
soliton model}

\author{\underline{N.Sawado}\footnote{sawado@ph.noda.tus.ac.jp},~N.Shiiki and S.Oryu}
\affiliation{Department of Physics, Tokyo University of Science, 2641, Yamazaki, 
Noda, Chiba 278-8510, Japan}

\begin{abstract}
We investigate higher baryon-number states in 
the Chiral Quark Soliton Model using the rational map ansatz for 
the background chiral fields. The soliton solutions are obtained 
self-consistently. We show that the baryon number density has 
point-like symmetries and the corresponding valence quark spectrum 
of the lowest energy exhibits approximate four-fold degenerate. 
Our results indicate the possibility of existence of the shell-like 
structure in the multi-baryonic system.   
\end{abstract}


\maketitle

 The Chiral Quark Soliton Model (CQSM) was developed in 1980's as an effective 
theory of QCD interpolating between the Constituent Quark Model and Skyrme 
Model \cite{diakonov,cqsm}. In the large $N_{c}$ limit, these models are identical 
\cite{manohar}. 

The CQSM is derived from the instanton 
liquid model of the QCD vacuum and incorporates the non-perturbative 
feature of the low-energy QCD, spontaneous chiral symmetry breaking.  
The vacuum functional is defined by;

\begin{eqnarray}
	{\cal Z} = \int {\cal D}\pi{\cal D}\psi{\cal D}\psi^{\dagger}\exp \left[ 
	i \int d^{4}x \, \bar{\psi} \left(i\!\!\not\!\partial
	- mU^{\gamma_{5}}\right) \psi \right]	 \label{vacuum_functional}
\end{eqnarray} 
where the SU(2) matrix
\begin{eqnarray}
	U^{\gamma_{5}}= \frac{1+\gamma_{5}}{2} U + \frac{1-\gamma_{5}}{2} U^{\dagger} 
	\,\,\,{\rm with} \,\,\,\,
	U=\exp \left( i \vec{\tau} \!\cdot\! \vec{\pi}/f_{\pi} \right) \nonumber
\end{eqnarray}
describes chiral fields, $\psi$ is quark fields and $m$ is the dynamical 
quark mass. $f_{\pi} $ is the pion decay constant and experimentally 
$f_{\pi} \sim 93 {\rm MeV}$. 
Since our concern is the tree-level pions and one-loop quarks according 
to the Hartree mean field approach, the kinetic term of the pion fields which 
gives a contribution to higher loops can be neglected. 
Due to the interaction between the valence quarks and the Dirac sea, 
soliton solutions appear as bound states of quarks in the background of self-consistent 
mean chiral field. $N_{c}$ valence quarks fill the each bound state to form a baryon. 
The baryon number is thus identified with the number of bound states filled by 
the valence quarks \cite{kahana}. 
For $B=1$ and $2$, the spherically symmetric soliton \cite{reinhardt,wakamatsu}
and the axially symmetric soliton \cite{sawado} were found respectively. Upon 
quantization, the intermediate states of nucleon and deuteron 
between the Constituent Quark Model and Skyrme Model were obtained. 

The vacuum functional in Eq.(\ref{vacuum_functional}) can be integrated 
over the quark fields to obtain the sea quark energy \cite{meissner2}

\begin{eqnarray}
	E_{sea}[U]=\frac{1}{4\sqrt{\pi}}N_{c}\int^{\infty}_{1/\Lambda^2}
	\frac{d\tau}{\tau^{3/2}}\left(\sum_{\nu}{\rm e}^{-\tau\epsilon_{\nu}^2}
	-\sum_{\nu}{\rm e}^{-\tau\epsilon^{(0)2}_{\nu}}\right)\,\, .
	\label{energy_sea} \label{enegy_sea}
\end{eqnarray}
where $\Lambda$ is cut-off parameter in the proper-time reguralization~\cite{schwinger} 
and $\epsilon_\nu,\epsilon_\nu^{(0)}$ are the eigenvalues of 
Dirac equation for a single quark with the vacuum sector defined by
\begin{eqnarray}
	\left(\frac{\vec{\alpha}\!\cdot\!\vec{\nabla}}{\it{i}}+\beta m 
	U^{\gamma_{5}}\right) |\nu\rangle = \epsilon_{\nu}|\nu\rangle\,\,,\,\,
	\left(\frac{\vec{\alpha}\!\cdot\!\vec{\nabla}}{\it{i}}+\beta m\right) 
	|\nu\rangle^{(0)} = \epsilon_{\nu}^{(0)}|\nu\rangle^{(0)}\,\,.
	\label{eigen_equation}
\end{eqnarray}

In the Hartree picture, the baryon states are the quarks occupying all 
negative Dirac sea and valence levels. Hence, if we define the total soliton energy 
$E_{total}$, the valence quark energy should be added;
 \begin{eqnarray}
	E_{total}[U]=N_{c}\sum_{i}E_{val}^{(i)}[U]+E_{sea}[U]\,\,. 
	\label{energy_total}
\end{eqnarray}
where $E_{val}^{i}$ is the valence quark contribution to the $i$ th baryon.

For higher winding number solitons, it is expected the solution to have 
point-like symmetries from the study of the Skyrme model \cite{braaten}. 
Therefore, we shall impose the same symmetries on the chiral fields 
using the rational map ansatz. According to the ansatz, the chiral field 
can be expressed as \cite{manton} 

\begin{eqnarray}
	U(r,z)=\exp \left( i f(r) \vec{n}_{R}\!\cdot\!\vec{\tau}\right) 
	\label{chiral_field} 
\end{eqnarray} 
where  
\begin{eqnarray*}
	\vec{n}_{R}=\frac{1}{1+|R(z)|^2}\left(\rm{2Re}[\it{R(z)}],
	\rm{2Im}[\it{R(z)}],\rm{1}-|\it{R(z)}|^2\right)
\end{eqnarray*}
and $R(z)$ is a rational map.

Rational maps are maps from $CP(1)$ to $CP(1)$ (equivalently, from $S^2$ 
to $S^2$) classified by winding number. It was shown in \cite{manton} 
that $B=N$ skyrmions can be well-approximated by rational maps with 
winding number $N$. The rational map with winding number $N$ possesses 
$(2N+1)$ complex parameters whose values can be determined by 
imposing the symmetry of the skyrmion. Thus, the rational maps 
for $B=2\sim7$ and $B=17$  take the form; 

\begin{eqnarray}
&&	R(z)=z^2~(B=2)\,\,,\,\,
	R(z)=\frac{\sqrt{3}iz^{2}-1}{z(z^{2}-\sqrt{3}i)}~(B=3)\,\,,\nonumber\\
&&    R(z)=\frac{z^4+2\sqrt{3}iz^2+1}{z^4-2\sqrt{3}iz^2+1}~(B=4)\,\,,\nonumber \\
&& 	R(z)=\frac{z(z^4+3.94z^2+3.07)}{3.07z^4-3.94z^2+1}~(B=5)\,\,,\,\,
R(z)=\frac{z(z^4-5)}{-5z^4+1}~(B=5^{*})\,\,,\nonumber \\
&&R(z)=\frac{z^4+0.16 i}{z^2(0.16z^4 i-1)}~(B=6)\,\,,\nonumber \\
&&R(z)=\frac{7/\sqrt{5}z^6-7z^4-7/\sqrt{5}z^2-1}{z(z^6+7/\sqrt{5}z^4+7z^2-7/\sqrt{5})}~(B=7)\,\,,\nonumber \\
&&R(z)=\frac{17z^{15}-187z^{10}+119z^5-1}{z^2(z^{15}+119z^{10}+187z^5+17)}~(B=17)\,\,,\,\,
	\label{rational_map}
\end{eqnarray}
where the complex coordinate $z$ on $CP(1)$ is identified with the 
polar coordinates on $S^2$ by $z=\tan (\theta /2){\rm e}^{i\varphi}$ 
via stereographic projection. Substituting (\ref{rational_map}) 
into (\ref{chiral_field}), one obtains the complete form of the chiral 
fields with appropriate symmetry and winding number. 
Since the chiral fields in (\ref{chiral_field}) are parameterized by the 
polar coordinates, one can apply the numerical technique developed 
for $B=1$ to find higher soliton solutions~\cite{wakamatsu}.  

Demanding that the total energy in (\ref{energy_total}) be stationary 
with respect to variation of the profile function $f(r)$,
\begin{eqnarray*}
	\frac{\delta}{\delta f(r)}E_{total}=0 \,\, ,
\end{eqnarray*}
yeilds the field equation 
\begin{eqnarray}
	S(r)\sin f(r)=P(r)\cos f(r) \label{field_equation}
\end{eqnarray}
where 
\begin{eqnarray}
&&S(r)=N_{c}\sum_{\nu}\bigl(n_\nu\theta(\epsilon_{\nu})+{\rm sign}(\epsilon_{\nu})
{\cal N}(\epsilon_{\nu})\bigr)\langle \nu|\gamma^{0}\delta(|x|-r)|\nu\rangle \\	
&&P(r)=N_{c}\sum_{\nu}\bigl(n_\nu\theta(\epsilon_{\nu})+{\rm sign}(\epsilon_{\nu})
{\cal N}(\epsilon_{\nu})\bigr)\langle \nu|i \gamma^{0}\gamma^{5}\vec{n}_{R}
\cdot\vec{\tau}\delta(|x|-r)|\nu\rangle\,\, .
\end{eqnarray}

The procedure to obtain the self-consistent solution of Eq.(\ref{field_equation}) 
is that $1)$ solve the eigenequation in (\ref{eigen_equation}) under an assumed 
initial profile function $f_{0}(r)$, $2)$ use the resultant eigenfunctions and 
eigenvalues to calculate $S(r)$ and $P(r)$, $3)$ solve 
Eq.(\ref{field_equation}) to obtain a new profile function, $4)$ repeat $1)-3)$ 
until the self-consistency is attained.

\begin{center}
\begin{figure}[hbt]
\includegraphics[height=11cm,width=15cm]{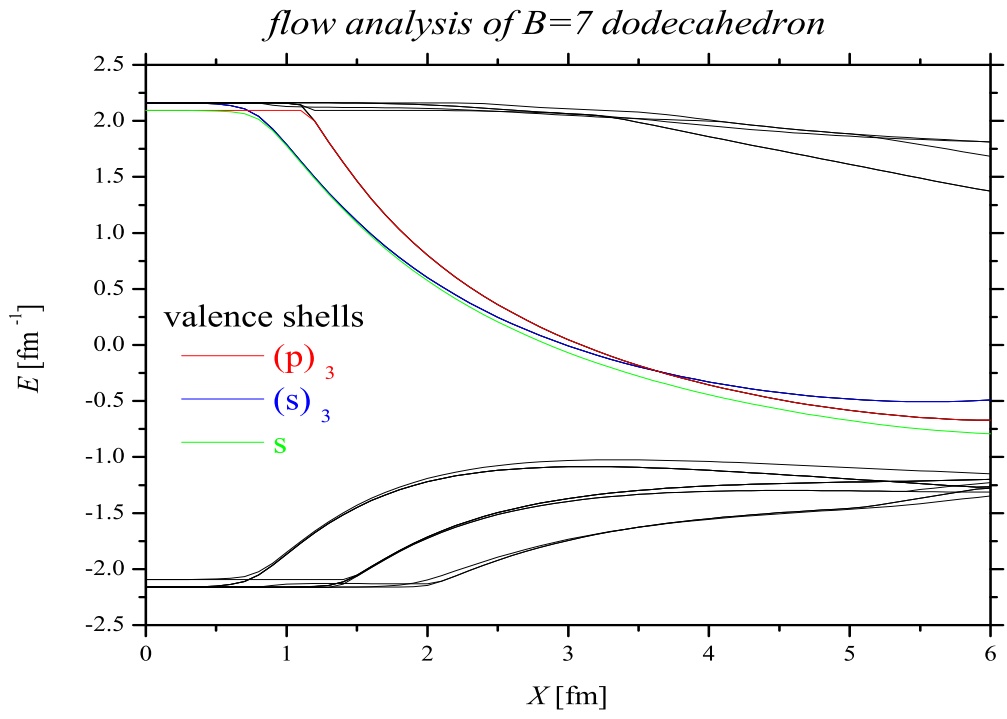}
\caption{\label{fig:spectrum7} Spectrum of the quark orbits for
$B=7$ dodecahedron. }
\end{figure}
\end{center}

Fig.~\ref{fig:spectrum3} $\sim$ Fig.~\ref{fig:spectrum7} shows the spectrums of the quark orbits for $B=3\sim7$ 
as a function of the soliton-size parameter $X$. 
In $B=3$ , the $P$ orbit diving into the negative energy region is triply degenerate. 
As discussed in~\cite{kahana}, baryon number of the soliton equals to the number 
of diving levels occupied by $N_{c}$ valence quarks. Thus putting $N_c=3$ valence 
quarks on each of the degenerate levels, one obtains the $B=3$ soliton solution.
Similarly, for higher $B$ than $3$, the same number of orbits as $B$ dive into 
negative energy region. 

\begin{figure}
\hspace{-2cm}
\begin{minipage}{3cm}
\includegraphics[height=4cm,width=4cm]{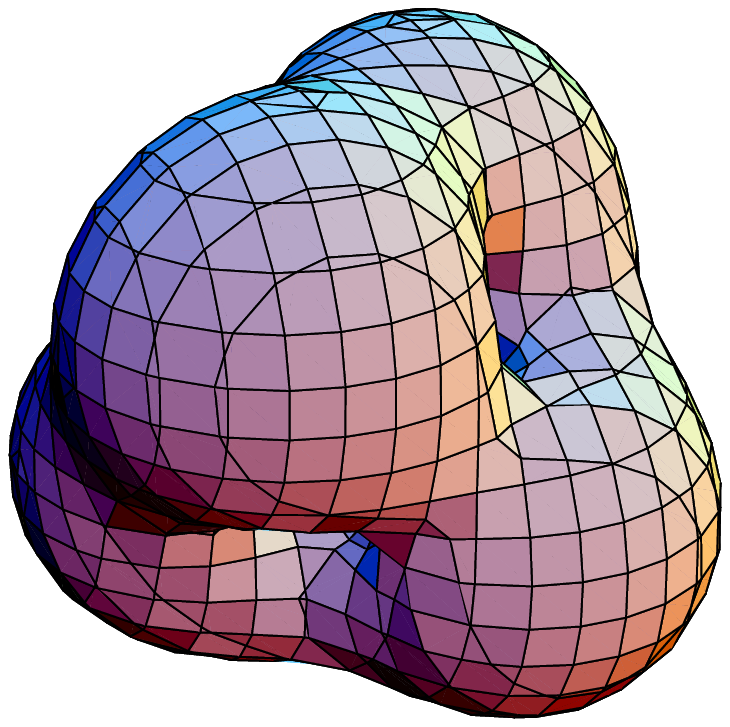}
\vspace{-5mm}

\hspace{-2cm}a
\end{minipage}
\hspace{2cm}
\begin{minipage}{3cm}
\includegraphics[height=4cm,width=4cm]{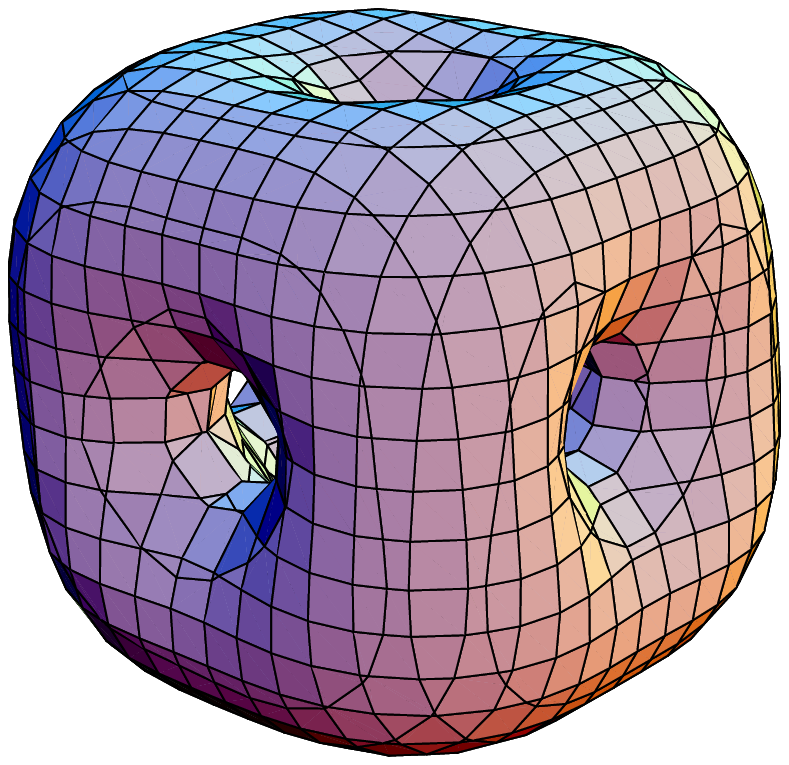}
\vspace{-3mm}

\hspace{-2cm}b
\end{minipage}
\hspace{2cm}
\begin{minipage}{3cm}
\includegraphics[height=4.5cm,width=4.5cm]{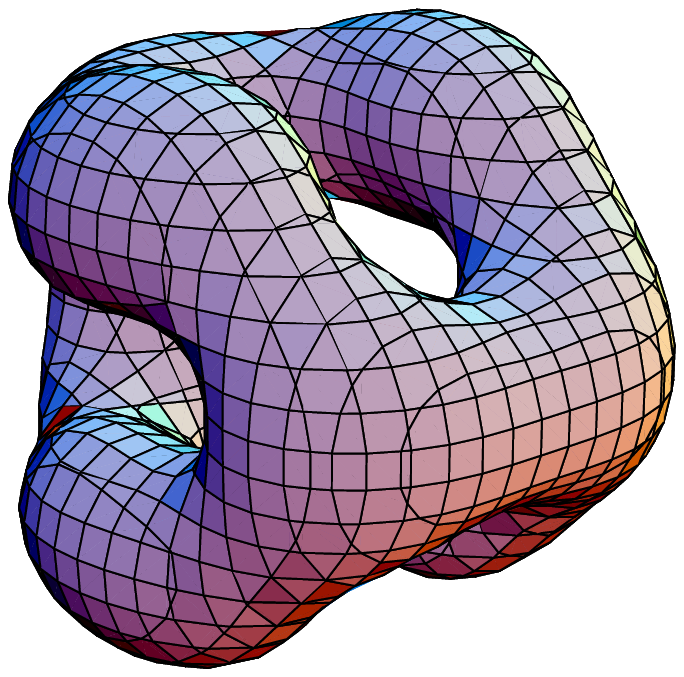}
\vspace{-10mm}

\hspace{-2cm}c
\end{minipage}\\
\hspace{-2cm}
\begin{minipage}{3cm}
\vspace{5mm}

\includegraphics[height=5cm,width=5cm]{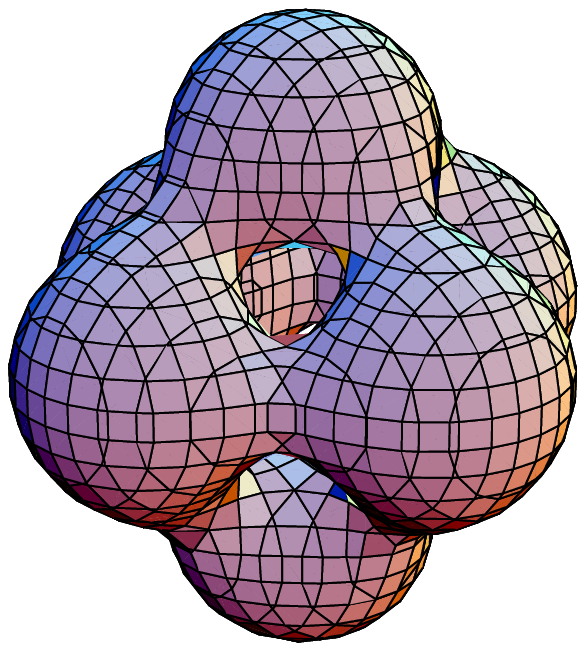}
\vspace{-10mm}

\hspace{-2cm}d
\end{minipage}
\hspace{2cm}
\begin{minipage}{3cm}
\vspace{5mm}

\includegraphics[height=5cm,width=5cm]{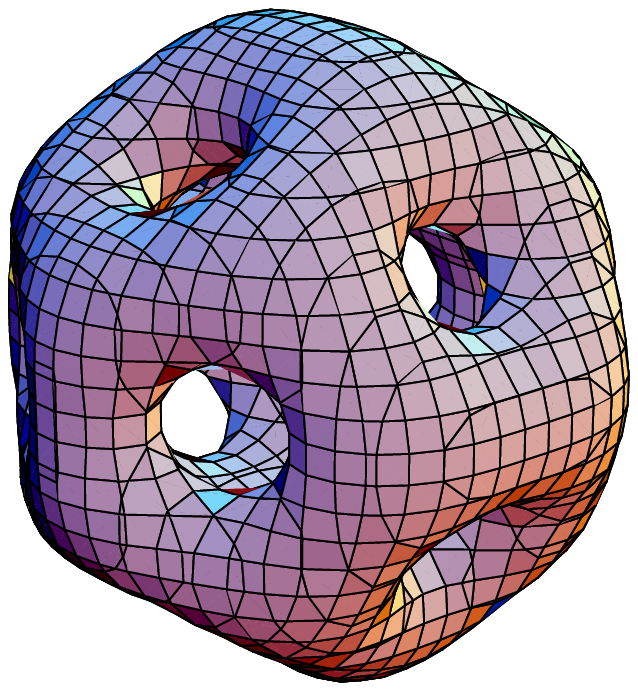}
\vspace{-11mm}

\hspace{-2cm}e
\end{minipage}
\caption{\label{fig:density} Surface of the baryon-number density:
(a)~$B=3$ tetrahedron, (b)~$B=4$ cube, (c)~$B=5$ with $D_{2d}$ symmetry, (d)~$B=5$ octahedron, 
(e)~$B=7$ dodedahedron.}
\end{figure}

\begin{center}
\begin{figure}
\includegraphics[height=11cm,width=15cm]{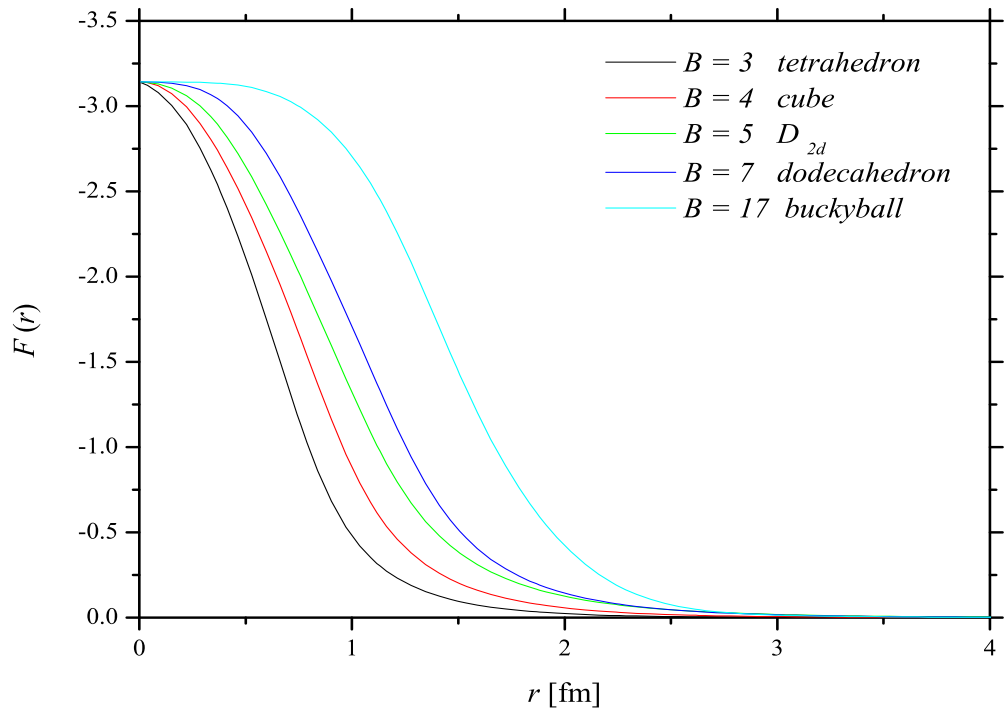}

\caption{\label{fig:profile} Profile functions $f(r)$ of the rational map ansatz. }
\end{figure}
\end{center}

Fig.~\ref{fig:density} and Fig.~\ref{fig:profile} show the 
corresponding baryon density and profile functions. 
This result is consistent with the result for skyrmions obtained 
by Braaten {\it et.al} \cite{braaten} and Houghton {\it et.al.}
~\cite{manton}.

\begin{figure}
\includegraphics[height=12cm,width=15cm]{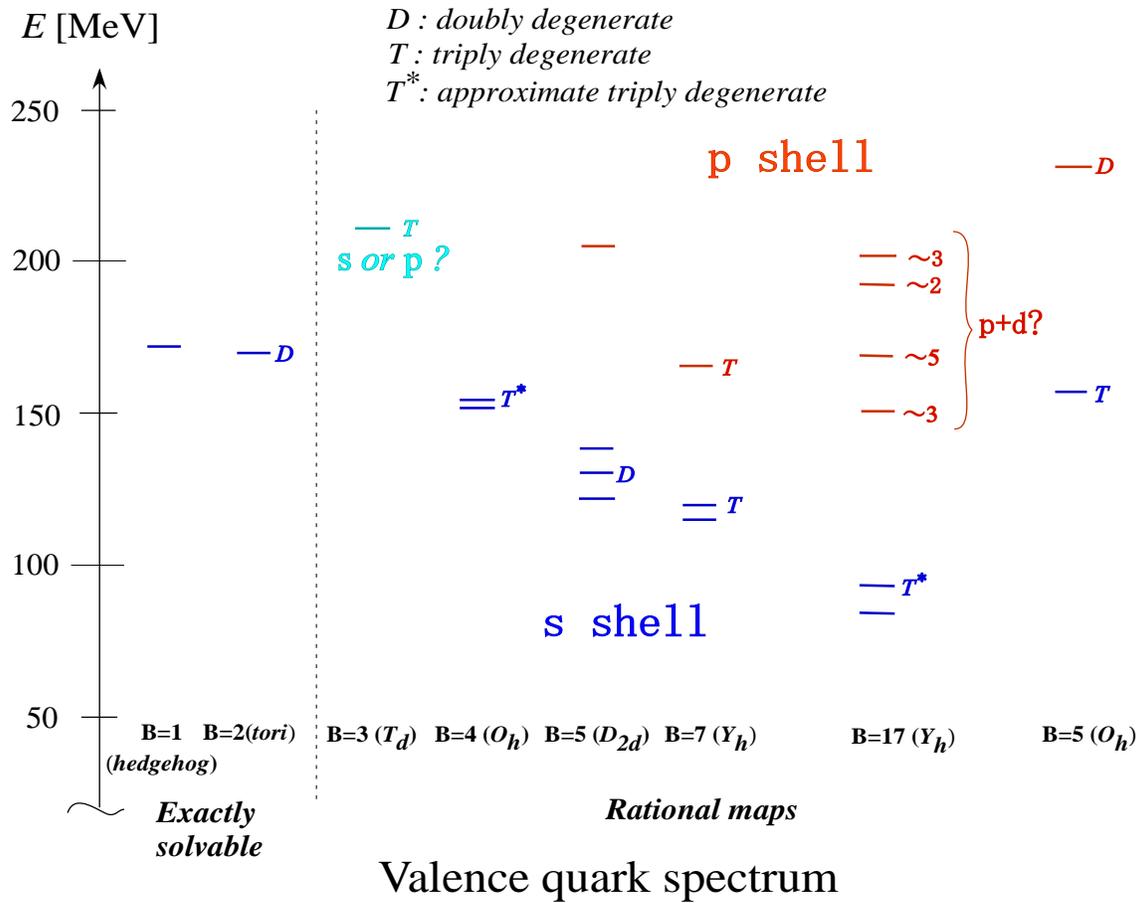}

\caption{\label{fig:spectra} Valence quark spectrums of $B=1\sim 7$ and $B=17.$}
\end{figure}

In Fig.~\ref{fig:spectra}, the valence quark spectra are shown 
for $B=1\sim 7$ and $B=17$. The approximate four-fold degeneracy 
can be observed for the lowest level. However, in the CQSM, 
a quark has only two quantum numbers: colour and grand spin defined 
by $\vec{K}=\vec{L}+\frac{1}{2}\vec{\sigma}+\frac{1}{2}\vec{\tau}$, 
which means the lowest level~($K=0$) should show no degeneracy.
At present, all we can say is that if one employs point-like symmetries 
for the background chiral fields, one obtains the larger degeneracy of 
a single quark states than the spherical symmetric case. 
Consequently, the number of the states can be reduced and the large shell gaps emerge. 
These results indicate the possibility of the existence of the lowest energy 
shell filled up with four quarks and also, these may be the reason why 
${}^4He$ is stable compared with other nuclei.   

\begin{center}
	{\bf Acknowledgements}
\end{center}
We are grateful to N.S.Manton for suggesting us this subject and 
the possibility of the shell structure for quarks.

\newpage
\begin{table*}[hbt]
\caption{\label{tab:helement} Spectrums of the valence, vacuum sea and their sum
~(in MeV). $\bar{M}$ is defined as $\bar{M}\equiv M/B M_{B=1}$ which means the 
energy per baryon.
The results for $\bar{M}$ in Skyrme model calculations~\cite{manton} are 
also quoted.}
\begin{ruledtabular}
\begin{tabular}{c|ccccccccccc}
$B$  & \multicolumn{7}{c}{Valence}        & Sea & Total & $\bar{M}$~(Skyrme)\\ \hline
1    & 173 &     &     &     &     &  &  & 674  & 1192 & 1.00~(1.00)\\
3    & 210 & 210 & 210 &     &     &  &  & 1554 & 3444 & 0.96~(0.96)\\ 
4    & 153 & 156 & 156 & 156 &     &  &  & 2544 & 4407 & 0.92~(0.92)\\
5    & 123 & 131 & 131 & 139 & 210 &  &  & 3265 & 5467 & 0.92~(0.93)\\ 
$5^{*}$& 157 & 157 & 157 & 232&232 &  &  & 2874 & 5680 & 0.95~(1.00)\\
7    & 115 & 120 & 120 & 120 & 166 & 166 & 166 & 4554 & 7478 & 0.90~(0.90)\\ 
\end{tabular}
\end{ruledtabular}
\end{table*}

\newpage
\begin{center}
\begin{figure}[hbt]
\includegraphics[height=11cm,width=15cm]{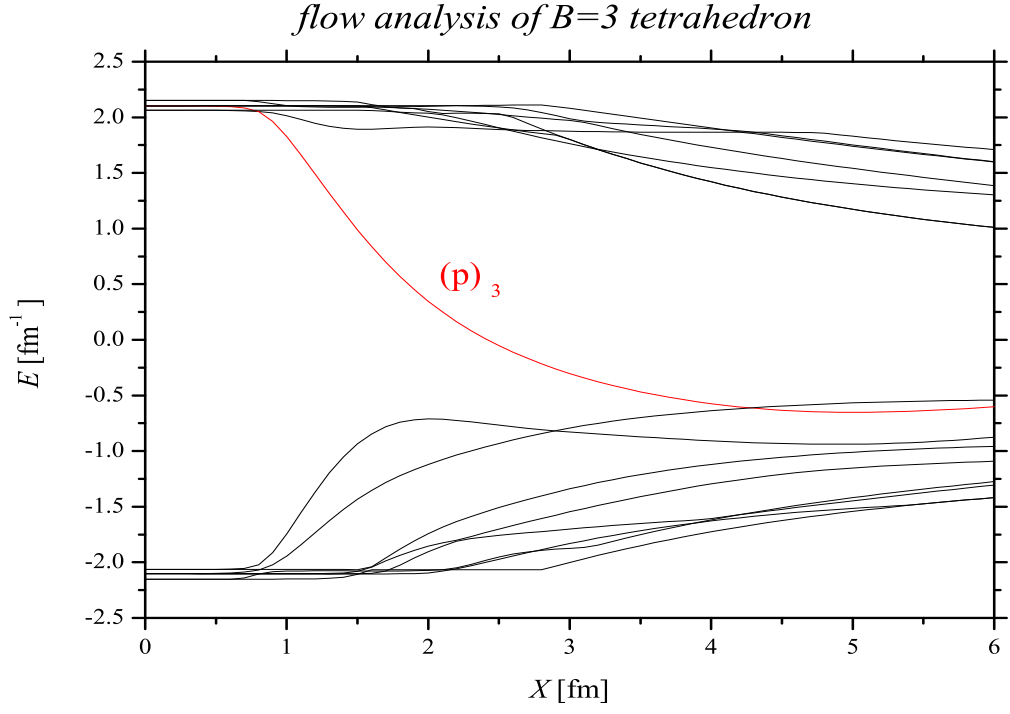}
\caption{\label{fig:spectrum3} Spectrum of the quark orbits for $B=3$ are 
illustrated as a function of the ``soliton size'' $X$. 
Trial profile function is defined by $f(r)=-\pi+\pi r/ X$ for $X \le r$ and $f(r)=0$ for $X > r$.
}
\end{figure}

\newpage
\begin{figure}[hbt]
\includegraphics[height=11cm,width=15cm]{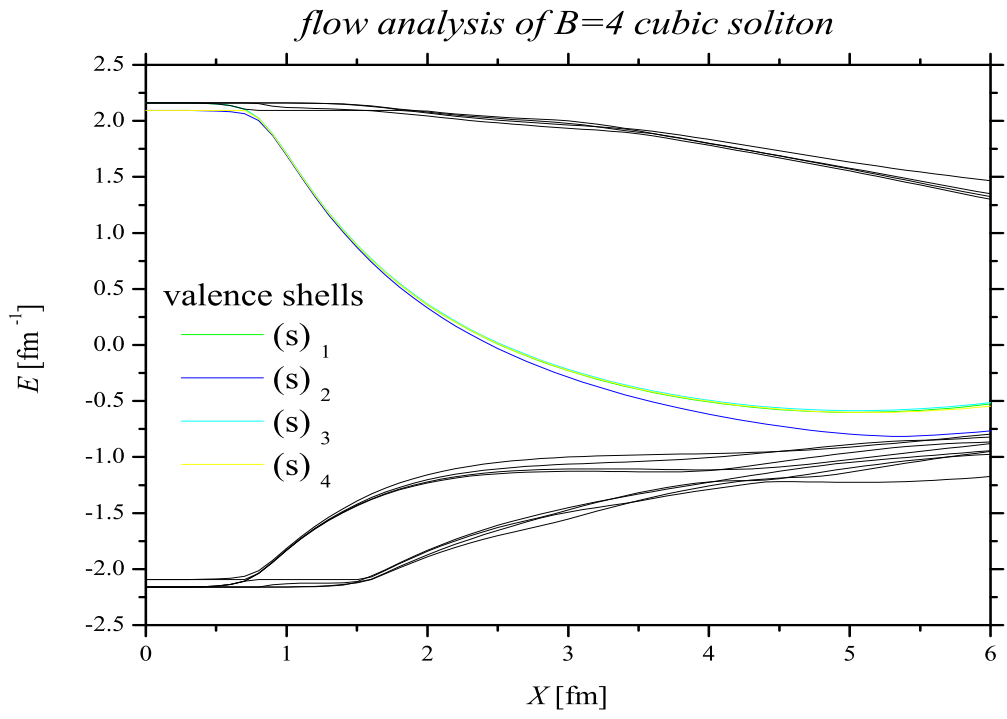}
\caption{\label{fig:spectrum4} Spectrum of the quark orbits for $B=4$. }

\end{figure}
\end{center}

\newpage
\begin{center}
\begin{figure}[hbt]
\includegraphics[height=11cm,width=15cm]{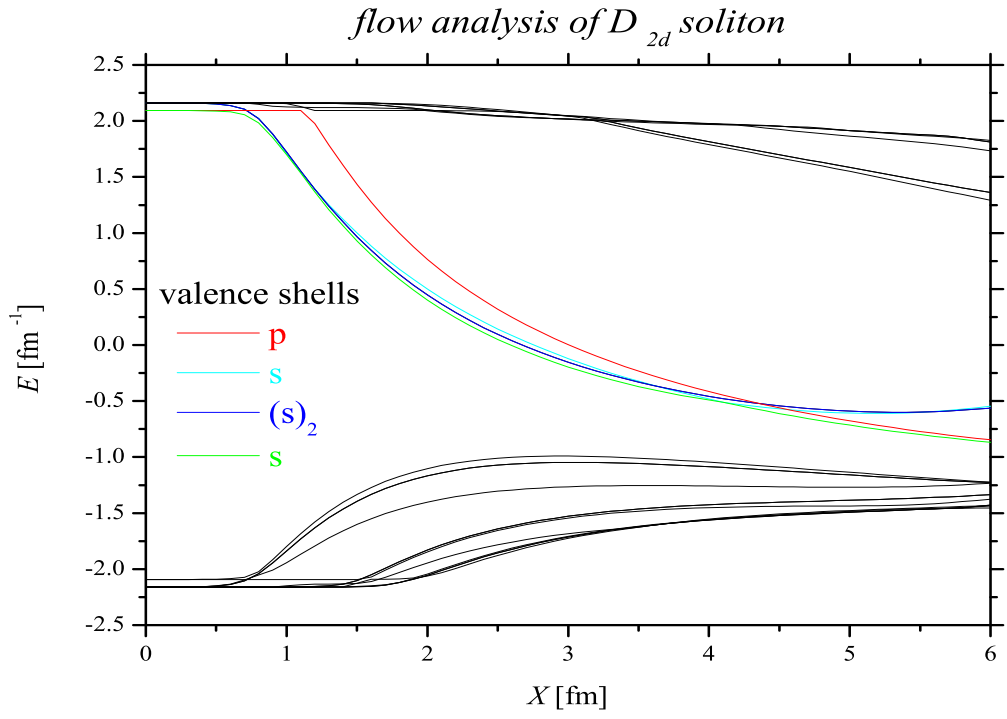}
\caption{\label{fig:spectrum5} Spectrum of the quark orbits for 
$B=5$ $D_{2d}$ symmetry.}
\end{figure}
\end{center}

\newpage
\begin{center}
\begin{figure}[hbt]
\includegraphics[height=11cm,width=15cm]{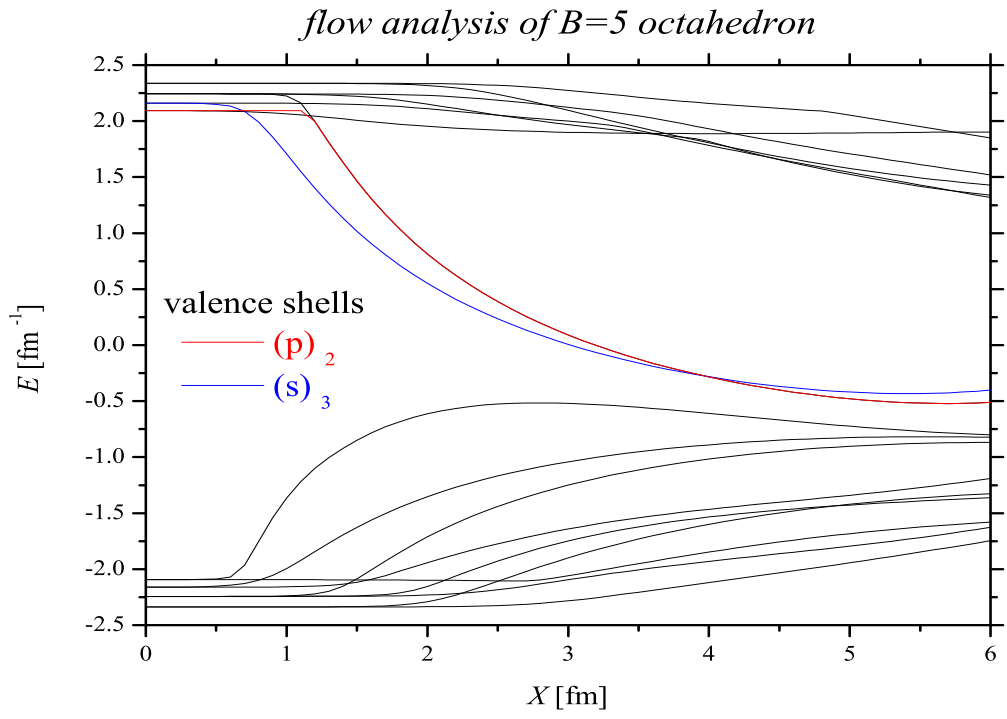}
\caption{\label{fig:spectrum5'} Spectrum of the quark orbits for $B=5$ octahedron.}
\end{figure}
\end{center}

\newpage

\newpage

\newpage

\newpage

\end{document}